\documentclass[journal,onecolumn,a4paper]{IEEEtran_nssmic}

\usepackage{cite}
\usepackage{graphicx}
\usepackage{stfloats}
\usepackage{amsmath}

\usepackage[latin1]{inputenc}

\newcommand{\kas}{KASCADE\ }

\newcommand{\kg}{KASCADE-Grande\ }
\newcommand{\kgp}{KASCADE-Grande}

\begin{document}
%
% paper title
\title{\rightline{\normalsize SI-HEP-2005-11}\rightline{}
A FADC-based Data Acquisition System for the KASCADE-Grande Experiment}%

\author{W. Walkowiak
  \thanks{W. Walkowiak was the presenter at the IEEE~2004 Nuclear Science
  Symposium and is the corresponding author 
  (e-mail: walkowiak@hep.physik.uni-siegen.de).}% 
  ,
  T.~Antoni,
  W.D.~Apel, 
  F.~Badea,
  K.~Bekk,
  A.~Bercuci,
  M.~Bertaina,
  H.~Bl\"umer, 
  H.~Bozdog,
  I.M.~Brancus,
  M.~Br\"uggemann,
  P.~Buchholz,
  C.~B\"uttner,
  A.~Chiavassa,
  K.~Daumiller,
  F.~di~Pierro,
  P.~Doll,
  R.~Engel,
  J.~Engler,
  F.~Fe{\ss}ler,
  P.L.~Ghia,
  H.J.~Gils,
  R.~Glasstetter,
  A.~Haungs,
  D.~Heck,
  J.R.~H\"orandel,
  K.-H.~Kampert,
  H.O.~Klages,
  Y.~Kolotaev,
  G.~Maier,
  H.J.~Mathes,
  H.J.~Mayer,
  J.~Milke,
  B.~Mitrica,
  C.~Morello,
  M.~M\"uller,
  G.~Navarra,
  R.~Obenland,
  J.~Oehlschl\"ager,
  S.~Ostapchenko,
  S.~Over,
  M.~Petcu,
  S.~Plewnia,
  H.~Rebel,
  A.~Risse,
  M.~Roth,
  H.~Schieler,
  J.~Scholz,
  M.~St\"umpert,
  T.~Thouw,
  G.~Toma,
  G.C.~Trinchero,
  H.~Ulrich,
  S.~Valchierotti,
  J.~van~Buren,
  A.~Weindl,
  J.~Wochele,
  J.~Zabierowski,
  S.~Zagromski, and 
  D.~Zimmermann\\[2ex] 
  \thanks{ % c
    A. Bercuci, I. M. Brancus, B. Mitrica, M. Petcu and G. Toma 
    are with National Institute of Physics and Nuclear Engineering,
    7690~Bucharest, Romania.}
  \thanks{ % h
    A. Risse and J. Zabierowski 
    are with Soltan Institute for Nuclear Studies, 90950~Lodz, 
    Poland.}
  \thanks{ % a
    T. Antoni, H. Bl\"umer, C. B\"uttner, J. R. H\"orandel, M. Roth
    and M. St\"umpert 
    are with Institut f\"ur Experimentelle Kernphysik,
    Universit\"at Karlsruhe, 76021 Karlsruhe, Germany.}
  \thanks{ % b
    W. D. Apel, F. Badea, K. Bekk, H. Bl\"umer, H. Bozdog, 
    K. Daumiller,  P. Doll, R. Engel, F. Fe{\ss}ler, H. J. Gils, 
    A. Haungs, D. Heck, H. O. Klages, G. Maier, H. J. Mathes,
    H. J. Mayer, J. Milke,  M. M\"uller, R. Obenland, J. Oehlschl\"ager,
    S. Ostapchenko, S. Plewnia, H. Rebel, H. Schieler, J. Scholz, 
    T. Thouw, H. Ulrich, J. van Buren, A. Weindl, J. Wochele
    and S. Zagromski
    are with Institut\ f\"ur Kernphysik, Forschungszentrum Karlsruhe,
    76021~Karlsruhe, Germany.}
  \thanks{ %e 
    M. Br\"uggemann, P. Buchholz, Y. Kolotaev, S. Over,  
    W. Walkowiak and D. Zimmermann
    are with Fachbereich Physik, Universit\"at Siegen, 57068 Siegen, 
    Germany.}
  \thanks{ % d
    M. Bertaina, A. Chiavassa, F. di Pierro, G. Navarra
    and S. Valchierotti 
    are with Dipartimento di Fisica Generale dell'Universit{\`a},
    10125 Torino, Italy.}
  \thanks{ % f 
    P. L. Ghia, C. Morello and G. C. Trinchero 
    are with Istituto di Fisica dello Spazio Interplanetario, CNR, 
    10133 Torino, Italy.}
  \thanks{ % g
    R. Glasstetter and K.-H. Kampert 
    are with Fachbereich Physik, Universit\"at Wuppertal, 42097
    Wuppertal, Germany.}
  \thanks{
    F. A. Badea is on leave of absence from Nat.\ Inst.\ of Phys.\ and 
    Nucl.\ Engineering, Bucharest, Romania.}
  \thanks{
    G. Maier is now at University Leeds, LS2~9JT~Leeds, United Kingdom.}
  \thanks{
    S. Ostapchenko is on leave of absence from Moscow State University, 
    119899~Moscow, Russia.}
}%

% make the title area
\maketitle

\begin{abstract}
We present the design and first test results 
of a new FADC-based data acquisition (DAQ)
system for the Grande array of the \kg experiment.  
The original \kas experiment at the Forschungszentrum Karlsruhe,
Germany, has been extended by 37~detector stations of the former
EAS-TOP experiment (Grande array) 
to provide sensitivity to energies of primary
particles from the cosmos of up to $\mathbf{10^{18}}$\:eV.  
The new FADC-based DAQ system will improve the quality of
the data taken by the Grande array by digitizing
the scintillator signals with a 250\:MHz sampling rate. 

The signals of each of the 37 detector stations are continuously
recorded using interleaved 12-bit Flash Analog-to-Digital-Converters
(FADCs) located on custom-made digitizer boards at each station. 
The digitizer boards feature a self-triggering mechanism, initiating the
data transmission to the Grande DAQ station using programmable
thresholds, a timestamp mechanism, and optical data transmission.  The
control logic is implemented using Field Progammable Gate Arrays
(FPGAs). 

Five optical receiver and temporary storage modules receive 
the data from up to eight stations each 
(at an approximate rate of 2.5\:MB/s per station) 
and transfer them into the memory of one of five PCs via a
customized PCI interface card.  Running on a master PC, the data 
acquisition software searches for coincidences in the timestamps, 
incorporates triggers generated by the other KASCADE-Grande components,
and builds air shower events from the data of the individual stations.  
Completed Grande events are sent to the central DAQ of
KASCADE-Grande at an approximate rate of a few air shower 
events per second.

Two Grande stations have been equipped with the FADC-based data
acquisition system and first data are shown.  

\end{abstract}


\section{Introduction}

\subsection{The \kg Experiment}

The  \kg experiment~\cite{kascadegrandeNIM2004}
is an astroparticle physics experiment
located at the site of the Forschungszentrum Karlsruhe, Germany,
at 110\:m above sea level
comprising a large collection area of about 0.5\:km$^2$.
\kg has been designed to study extensive air showers induced by 
primary particles in the energy range of $E_0 = 10^{14}$ to $10^{18}$\:eV.  
The major goal of the \kg experiment is the determination of the chemical 
composition of the primary particles in the energy range of
the so-called ``knee'' in the cosmic-ray energy
spectrum. Specifically, high quality data are
needed to clearly identify the expected knee in the iron spectrum 
at about $10^{17}$\:eV.
The measurement of the iron spectrum in this region 
will allow \kg to e.g. 
distinguish between different models for acceleration mechanisms of
high-energy particles in the cosmos. 

\begin{figure}[t]
  \centering
  \vspace*{-0.2cm}%
  \includegraphics[width=0.40\textwidth]{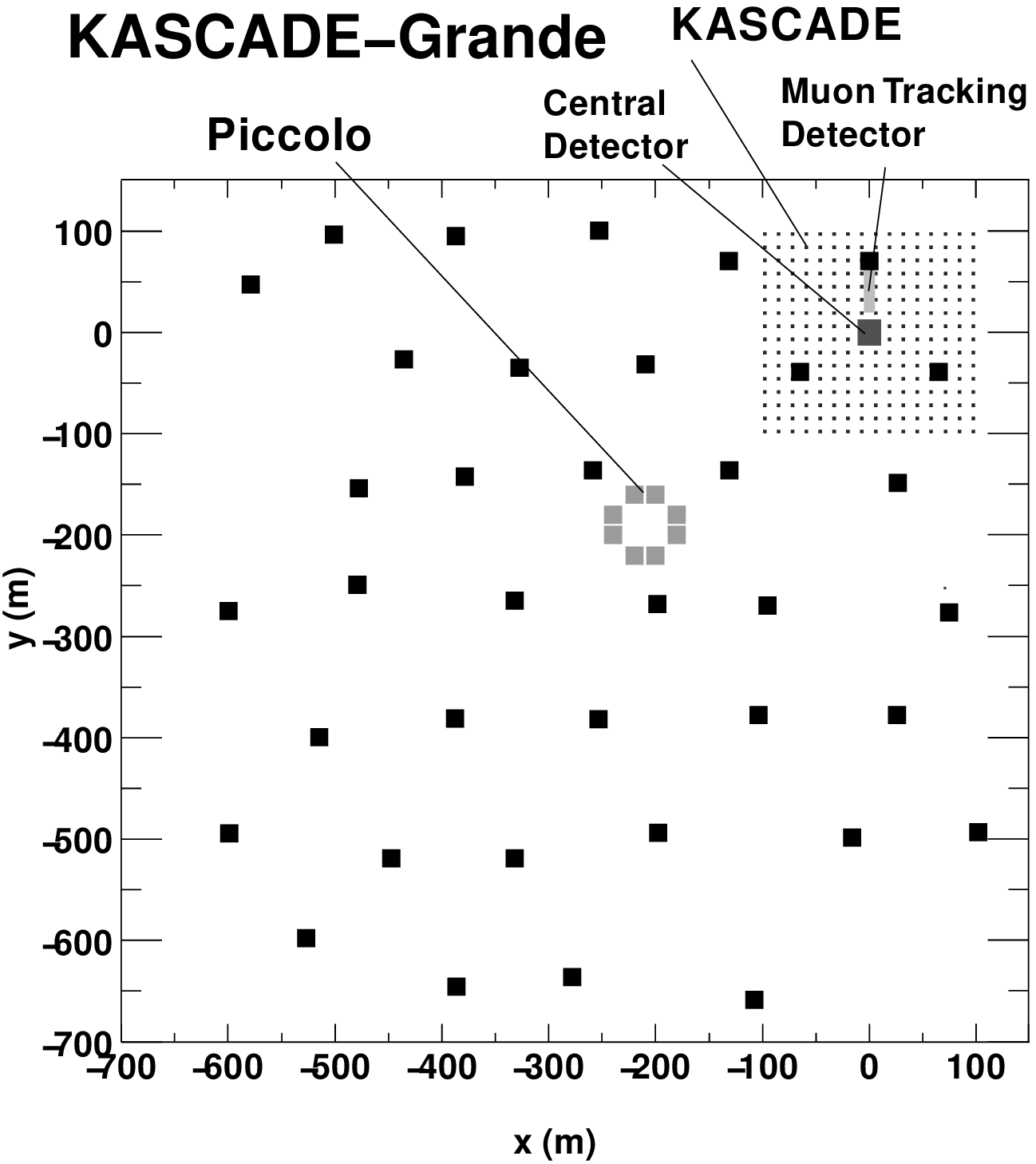}
  \vspace*{-0.2cm}%
  \caption{Layout of the KASCADE-Grande experiment 
    \cite{kascadegrandeNIM2004}.}
  \label{fig1}
\end{figure}

\begin{figure}[t]
  \centering
  \vspace*{-0.2cm}
  \includegraphics[width=0.50\textwidth]{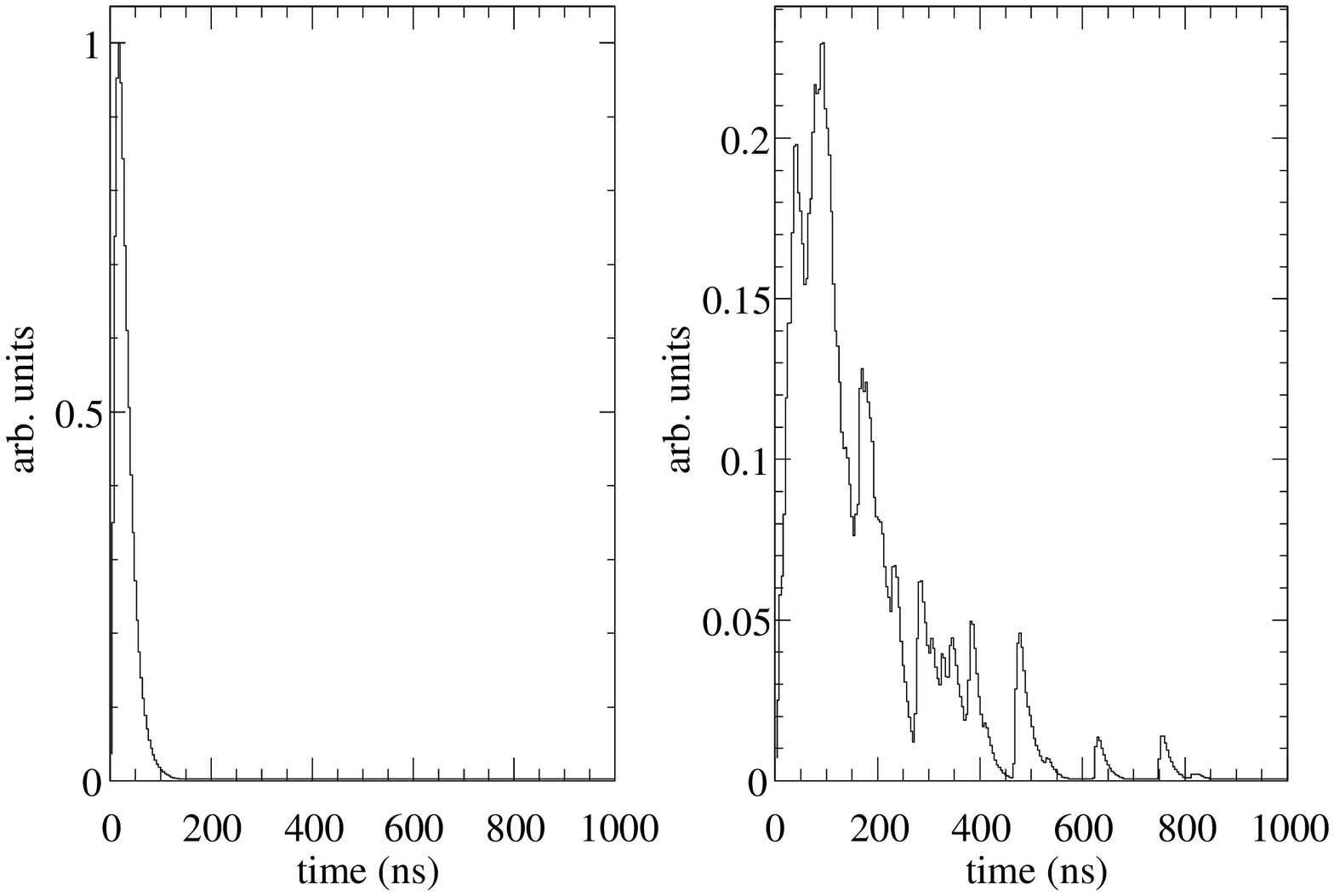}
  \vspace*{-0.7cm}
  \caption{Simulated signal shape in a single station for
    a $10^{18}$\:eV iron induced shower at 66.5\:m (left) 
    and 493.8\:m (right) distance from the shower core.}
  \label{fig2}
  \vspace*{-0.2cm}
\end{figure}

\begin{figure*}[!b]
  \centering
  \vspace*{-0.2cm}
  \includegraphics[width=0.95\textwidth]{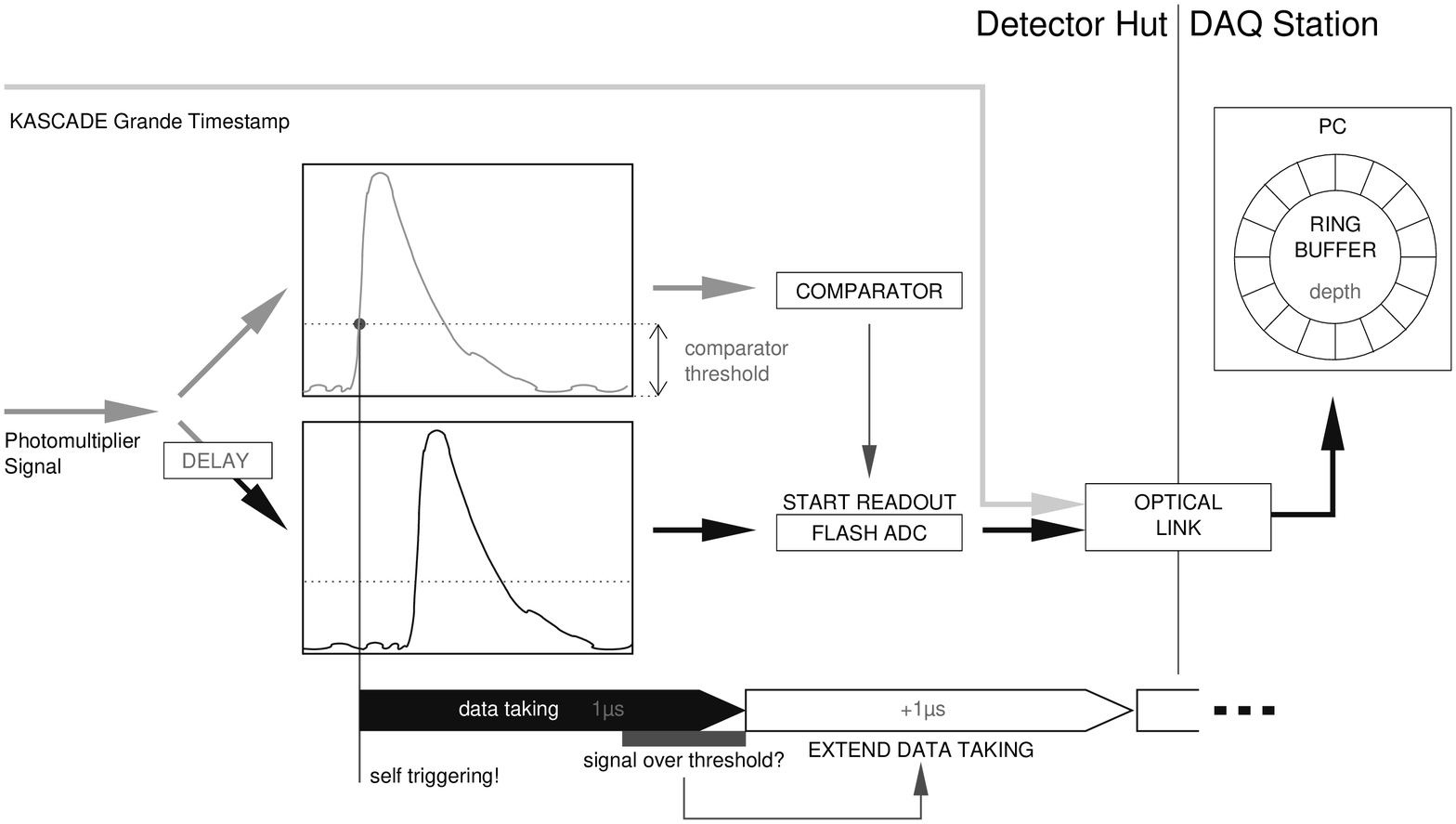}
  \vspace*{-0.2cm}
  \caption{Concept of the FADC-based data acquisition system.~\cite{over2004}}
  \label{fig3}
\end{figure*}

\kgp~\cite{kascadegrandeNIM2004} (fig.~\ref{fig1}) consists of the 
KASCADE experiment~\cite{kascadeNIM2003}, the 
Grande array with 37 stations of
10\:m$^2$ scintillation counters each -- re-used from the former EAS-TOP
experiment~\cite{Aglietta:1993gq} -- at an average mutual distance of
approximately 137\:m, and the trigger scintillator array Piccolo. 
The Piccolo array is 
placed between the centers of the Grande and KASCADE arrays and provides
a common trigger to both, resulting in full detection efficiency for $E_0 
\stackrel{>}{~} 10^{16}$\:eV.
The KASCADE experiment consists of a square grid  ($200\times200$\:m$^2$)
of 252 electron and muon detectors, an underground muon tracking
detector for reconstructing the muon direction with an accuracy
of $\sigma \approx 0.35^\circ$, and the central detector, which is
mainly used for hadron measurements \cite{kascadeCaloNIM1999}.  

\begin{figure*}[!t]
  \centering
  \vspace*{-0.2cm}
  \includegraphics[width=0.90\textwidth]{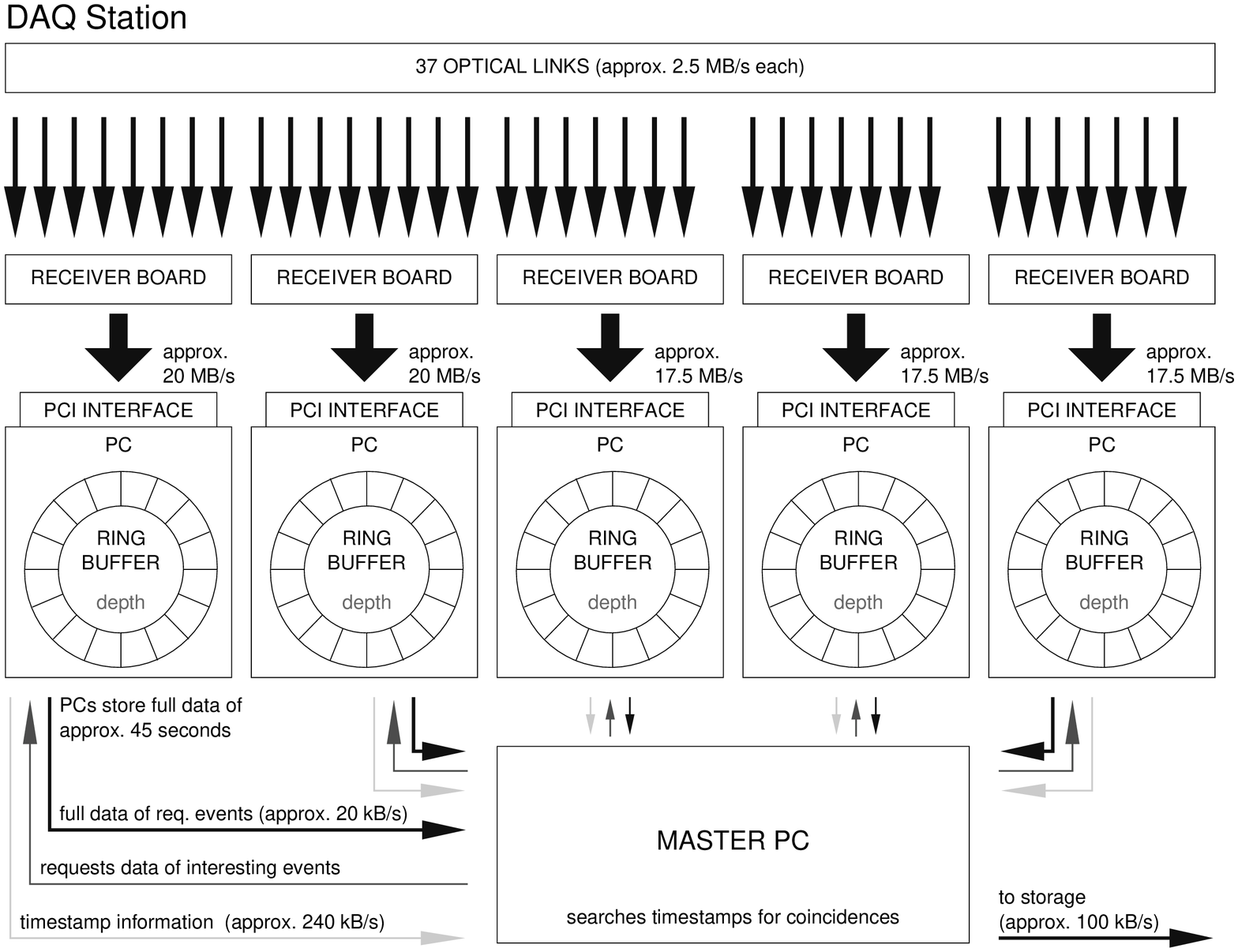}
  \vspace*{-0.2cm}
  \caption{Data collection and event building in the DAQ station.~\cite{over2004}}
  \label{fig4}
\end{figure*}

\begin{figure}[t]
  \centering
  \vspace*{-0.2cm}
  \includegraphics[width=0.45\textwidth]{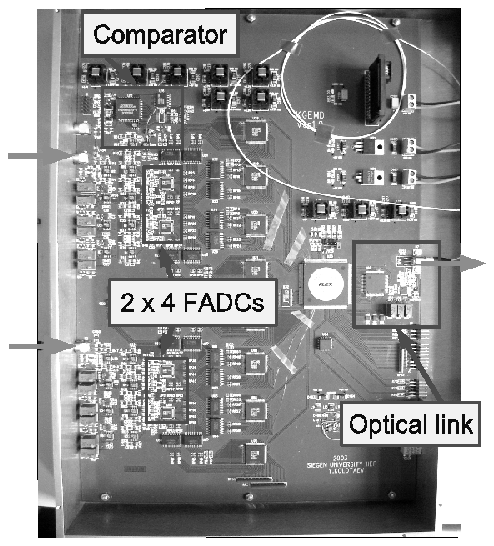}
  \caption{Photograph of the digitizer board (KGEMD).
  The high and low gain inputs are indicated on the left hand side.}
  \label{fig3a}
\end{figure}

\begin{figure}[b]
  \centering
  \vspace*{-0.2cm}
  \includegraphics[width=0.45\textwidth]{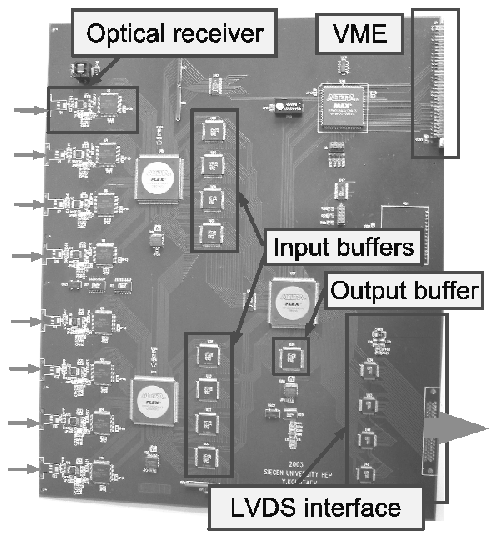}
%%  \vspace*{-0.2cm}
  \caption{Photograph of the receiver module board (KGEMS).  
    The optical inputs
    from up to eight detector stations are indicated on the left hand
    side.}
  \label{fig4a}
\end{figure}

\begin{figure}[t]
  \centering
  \vspace*{-0.5cm}
  \includegraphics[width=0.45\textwidth]{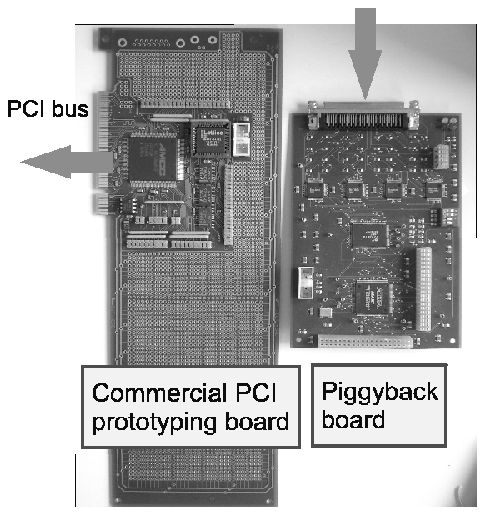}
  \caption{Photograph of the PCI interface card (KGEMP).  
    The LVDS input from the one receiver board servicing 
    up to eight detector stations is indicated in the upper right 
    corner.}
  \label{fig4b}
\end{figure}

\subsection{The Grande DAQ System}

Each
Grande station contains 16 ($80 \times 80$\:cm$^2$) plastic scintillator
plates read out by photomultipliers with low and high gain resulting
in a combined dynamic range of 0.3 to $30\,000$ vertical muons per station.  
In the current implementation the signals are amplified and then
shaped inside the individual
stations, transmitted to the central Grande DAQ station via 700\:m long 
shielded cables, and digitized by peak sensing ADCs, thus providing a
single value per station corresponding to the total energy deposit in
this station. For triggering purposes, the Grande array is electronically
subdivided into 18 overlapping hexagonal clusters of 7\:stations each.  For a
4-fold coincidence within each cluster a trigger rate of $\sim 0.3$\:Hz
per cluster and of $\sim 5.9$\:Hz for the total of the Grande array has been
measured~\cite{kascadegrandeKamp2003}.  The single muon rate per
station has been measured to be approximately 2.2\:kHz~\cite{over2004}.

In the new implementation, in order to record the full time-evolution 
of energy deposits in the Grande stations induced by air showers, a new
self-triggering, 
dead-time free FADC-based DAQ system with continuous 4\:ns-sampling
has been developed.  Using the
fine-grained sampling of the shower shapes we will gain additional
information about shower characteristics, e.g. about the particle
density development in the shower front or about the distance of a
specific Grande station with respect to the shower core, since the
expected pulse shapes for showers far from the shower core
clearly differ from those close to the shower core (fig.~\ref{fig2}).
The 250\:MHz sampling rate will provide a measurement of the shower
front arrival time with an accuracy of about 1\:ns or
less~\cite{andrei2004}.
 
\section{The Hardware of the FADC-based DAQ System}

The new self-triggering FADC-based DAQ system features high time
resolution of the signal shapes and an optical data transfer 
(fig.~\ref{fig3}).

The 
scintillator pulses of a Grande detector are digitized on the
digitizer board (KGEMD) by FADCs running in interleaved mode 
with an effective sampling rate of 250\:MHz. 

Fig.~\ref{fig4} depicts the part of the FADC data
acquisition system which is being installed in the Grande DAQ station.
The system consists of five optical receiver modules (KGEMS) which
serve up to eight detector stations each, five DAQ computers with
customized PCI interface cards (KGEMP) for temporary storage, 
and a master PC which is responsible for the event building.
The DAQ computers are connected to the master PC via a private 1000~MBit/s
ethernet.

\subsection{Event Digitization in the Grande Stations}
 
The KGEMD (fig.~\ref{fig3a})
consists of two times four 12-bit FADCs (one set of four per gain
range) running at 62.5\:MHz 
and Field-Programmable-Gate-Arrays (FPGAs) which provide the necessary
logic for the data management at readout and the generation of time
stamps from external 1\:Hz and 5\:MHz signals \cite{kascadeNIM2003}.
  The four FADCs per
channel are operated in interleaved mode, i.e. with a relative 
time offset of 4\:ns (corresponding to 90\:degree phase shifted sampling 
clocks), such that the effective sampling rate of 
250\:Msamples/s is obtained.  While the FADCs are continuously
digitizing the waveform,
their data are only transferred to buffering FIFOs once the adjustable 
comparator threshold is crossed.  Taking advantage of the internal
delay of the FADC, which corresponds to 7 clock cycles, the recorded 
signal waveform contains data for up to 112\:ns ahead of the threshold
crossing time.  The signal waveform has a typical length of about
1\:$\mu$s.  Time stamp information of several counters, which are
synchronized by external 1\:Hz and 5\:MHz signals,
and an identifier for the station
number are added to the data packet before it is transmitted to the
Grande DAQ station via an optical link (1300\:nm laser diode, distance
up to 800\:m, data transfer rate of 1.25\:GBaud).  The data packet 
for a single 1\:$\mu$s long event containing both, high and low gain data, 
consists of 512 16-bit words, i.e. a total of 1\:kByte data per event.
In case the input signal exeeds the threshold during the last
part (typically the last 200\:ns) of the previous 1\:$\mu$s period,
another 1\:$\mu$s long data packet is seamlessly added.

\subsection{Data Collection and Event Building in the DAQ station}

Fig.~\ref{fig4a} shows a photograph 
of the 9U~VME receiver module, KGEMS.  
In order to derandomize the incoming data, each of the eight optical
input channels (left side) is buffered by a FIFO, which can
hold up to 16 full single station event data packets.  These are 
multiplexed onto one output buffer using FPGAs
and passed on to the PCI interface card via a 32-bit wide LVDS link.  
The VME bus is used to upload the configuration and for debugging purposes.

The PCI interface (fig.~\ref{fig4b}) consists of a commercial PCI
prototyping board and a custom made piggy-back card, which receives the 
data via the LVDS link from the receiver board.  The PCI-interface
uses the Direct Memory Access (DMA) mechanism to write the data to a
ring buffer inside the DAQ computer.  While 128\:MB of the 1\:GB memory
of the DAQ computer have been reserved for the Linux operating system,
the major part (896\:MB) is used for incoming events. A custom-written
driver enables user space programs to access this ring buffer as
events become available~\cite{over2004}.
Data transfer rates to the PC's memory of 85\:MByte/s
have been achieved, while on average a rate of 20\:MByte/s is needed.
At this nominal rate the ring buffer memory may hold up to 45\:s of
event data.  In addition, the DAQ PCs have the capability to generate
high statistic histograms for any desired quantity 
based on single station events, such as distributions related
to the energy per event deposited in a single station.  The typical
shape of the latter provides input to an energy calibration of the
Grande detector stations (by comparison with the expected shape from
a Monte Carlo simulation).

The master PC only obtains the timestamp information from the 
DAQ PCs (approximate rate of 30\:kB/s) and scans it for coincidences.
It requests the corresponding data
from the DAQ computers for interesting events only.
After event building the events are put to 
mass storage and can be 
forwarded to the central DAQ of KASCADE-Grande at an approximate rate
of 100\:kByte/s.  The system allows for a flexible definition of the
selection criteria.  More sophisticated criteria than simple coincidences
in time may be implemented in the future.

\begin{figure}[t]
  \centering
  \vspace*{-0.2cm}
  \includegraphics[width=0.46\textwidth]{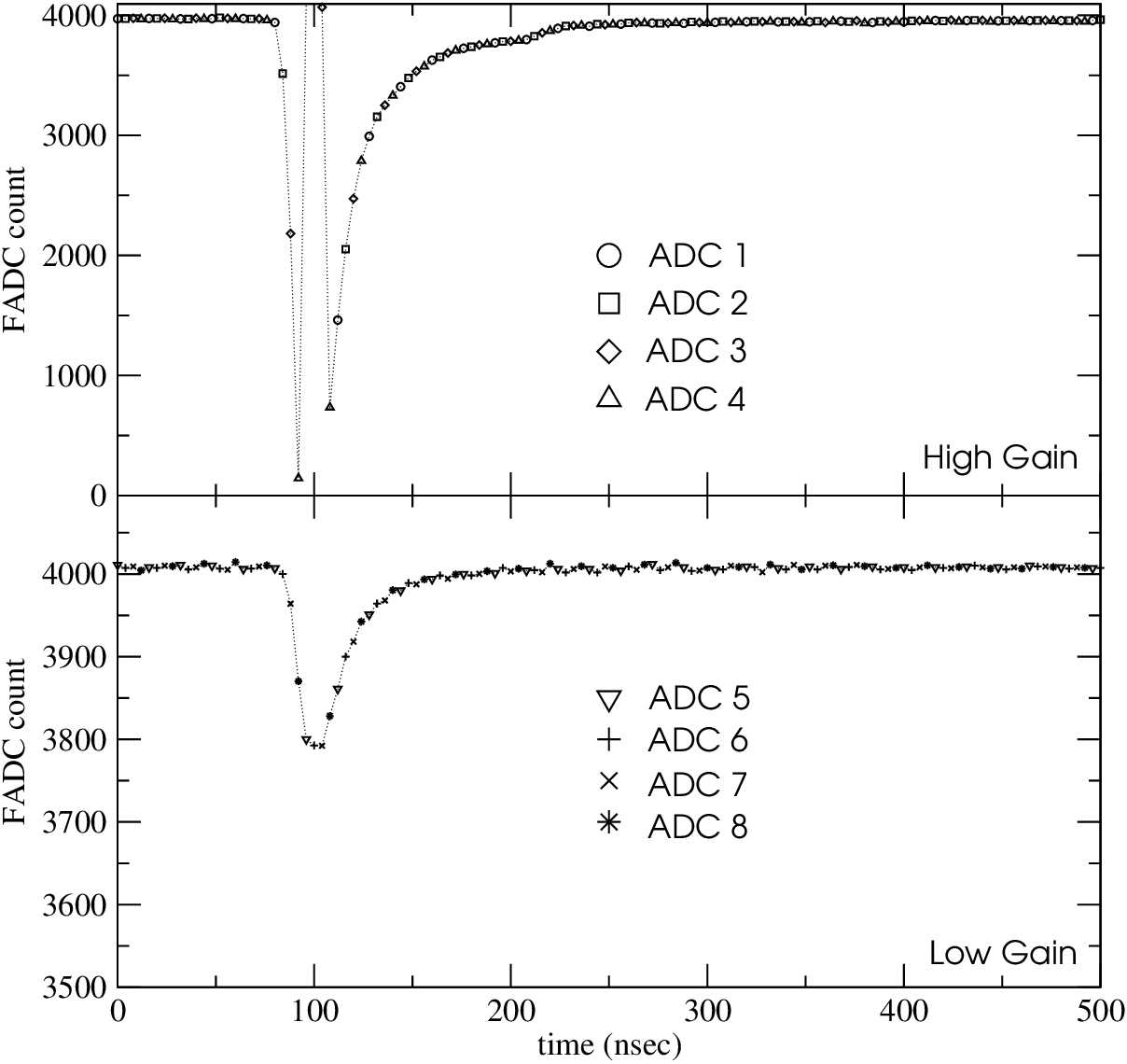}
  \vspace*{-0.2cm}
  \caption{Sample signal event in the high and low gain channels
    for one station.  The data of the 
    four interleaved FADCs per channel are marked by different symbols.}
  \label{fig5}
\end{figure}

\section{System installation} 

Recently, 
two Grande detector stations have been equipped with one
digitizer board each.  One receiver module servicing the corresponding
two optical links has been set up in conjunction with the necessary PC
and it's PCI-interface card.  Single station events have been recorded
at an approximate rate of 2.5\:kHz per station.  A sample event is
shown in (fig.~\ref{fig5}).  The different markers denote the data
taken by the four different FADCs of each channel, which have been
intercalibrated in order to account for slightly different gains and
offsets of the individual FADCs. This picture nicely demonstrates the need
for the high- and low-gain channels:  While the high gain channel is
saturated by the peak, there is still ample room for even taller
signal peaks in the low gain channel.

Production of the necessary components for the full system for all
37~detector stations is progressing.  The completion of the FADC DAQ
system is envisaged for 
%% the fall of 
2005.  Using the input from the
two detector stations already equipped, the event building software is
currently being implemented.

\section{Conclusions}

A self-triggering, dead-time free FADC-based DAQ system with 
a 4\:ns sampling and high resolution of 12 bits in two 
input gain ranges each has been designed 
for the \kg experiment and is being built.  
The system will provide \kg with the full pulse shape
information covering the entire thickness of the shower front. 
This additional information will greatly improve the data quality and 
prove valuable for the reconstruction of extensive air showers as new
(or improved) observables are identified.  

Recently, two Grande stations of \kg have been equipped with FADC-based 
digitizer boards and the DAQ chain up to the first DAQ computer has
been operated successfully.  
In 2005 data taking with the 
FADC system for all 37 detector stations is scheduled to commence. 

\newpage 

\section{Acknowledgements}

The authors
would like to thank the technical staff at Siegen and at the
Forschungszentrum Karls\-ruhe, 
who have been working on the design, production, 
testing and installation of the FADC system.  Their dedication made it
possible for us to begin with the installation of the
system during the summer of 2004.  

The \kg experiment is supported by the Federal Ministry of Education and
Research (BMBF) of Germany, the Polish State Committee for Scientific
Research (KBN grant for the years 2004 to 2006) and the Romanian
National Academy for Science, Research and Technology.

\bibliographystyle{IEEEtran}
\bibliography{IEEEabrv,IEEE2004_tns}

% that's all folks
\end{document}